\documentclass[twocolumn,twocolappendix]{aastex63}

\usepackage{color} 
\usepackage{float}
\usepackage{amsmath}
\usepackage{array}
\newcommand{\red}[1]{\textcolor{red}{#1}}

\newcommand{\rom}[1]{\MakeUppercase{\romannumeral #1}}
\newcolumntype{P}[1]{>{\centering\arraybackslash}p{#1}}

\shorttitle{Evidence for Gas Accretion}
\shortauthors{Luo et al.}

\begin{document}

\title{Evidence for the Accretion of Gas in Star-Forming Galaxies: High N/O Abundances in Regions of Anomalously-Low Metallicity}

\correspondingauthor{Yuanze Luo}
\email{yluo37@jhu.edu}

\author[0000-0002-0696-6952]{Yuanze Luo}
\affiliation{Department of Physics and Astronomy,
Johns Hopkins University,
Baltimore, MD 21218 USA}

\author[0000-0001-6670-6370]{Timothy Heckman}
\affiliation{Department of Physics and Astronomy,
Johns Hopkins University,
Baltimore, MD 21218 USA}

\author[0000-0003-4250-4437]{Hsiang-Chih Hwang}
\affiliation{Department of Physics and Astronomy,
Johns Hopkins University,
Baltimore, MD 21218 USA}

\author[0000-0001-7883-8434]{Kate Rowlands}
\affiliation{Department of Physics and Astronomy,
Johns Hopkins University,
Baltimore, MD 21218 USA}
\affiliation{AURA for ESA, Space Telescope Science Institute,
3700 San Martin Drive, Baltimore, MD 21218, USA}

\author[0000-0003-1888-6578]{Laura S\'{a}nchez-Menguiano}
\affiliation{Instituto de Astrof\'isica de Canarias, 38205 La Laguna, Tenerife, Spain} 
\affiliation{Departamento de Astrof\'isica, Universidad de La Laguna, Spain}
\affiliation{European Southern Observatory, Karl-Schwarzschild-Str. 2, Garching bei Munchen, D-85748, Germany}

\author[0000-0002-1321-1320 ]{Rog\'{e}rio Riffel}
\affiliation{Departamento de Astronomia, Universidade Federal do Rio Grande do Sul, IF, CP 15051, Porto Alegre 91501-970, RS, Brazil}
\affiliation{Laborat\'orio Interinstitucional de e-Astronomia - LIneA, Rua Gal. Jos\'e Cristino 77, Rio de Janeiro, RJ - 20921-400, Brazil}

\author[0000-0002-3601-133X]{Dmitry Bizyaev}
\affiliation{Apache Point Observatory and New Mexico State University, P.O. Box 59, Sunspot, NM, 88349-0059, USA}
\affiliation{Sternberg Astronomical Institute, Moscow State University, Moscow, Russia}

\author[0000-0001-8085-5890]{Brett H. Andrews}
\affiliation{Department of Physics and Astronomy and Pittsburgh Particle Physics, Astrophysics, and Cosmology Center (PITTPACC), University of Pittsburgh, 3941 O'Hara Street, Pittsburgh, PA 15260, USA}

\author[0000-0003-3526-5052]{Jos\'{e} G. Fernández-Trincado}
\affiliation{Instituto de Astronom\'{i}a y Ciencias Planetarias de Atacama, Universidad de Atacama, Copayapu 485, Copiap\'{o}, Chile}

\author[0000-0002-7339-3170]{Niv Drory}
\affiliation{McDonald Observatory, The University of Texas at Austin, 1 University Station, Austin, TX 78712, USA}

\author[0000-0003-1123-6003]{Jorge S\'{a}nchez Almeida}
\affiliation{Instituto de Astrofısica de Canarias, 38205 La Laguna, Tenerife, Spain}

\author[0000-0002-4985-3819]{Roberto Maiolino}
\affiliation{Cavendish Laboratory, University of Cambridge, 19 J. J. Thomson Avenue, Cambridge CB3 0HE, United Kingdom}
\affiliation{Kavli Institute for Cosmology, Cambridge, University of Cambridge, Madingley Road, Cambridge, CB3 0HA, United Kingdom}

\author{Richard R. Lane}
\affiliation{Instituto de Astronom\'{i}a y Ciencias Planetarias de Atacama, Universidad de Atacama, Copayapu 485, Copiap\'{o}, Chile}

\author[0000-0002-0789-2326]{Maria Argudo-Fern\'{a}ndez}
\affiliation{Instituto de F\'{i}sica, Pontificia Universidad Cat\'{o}lica de Valpara\'{i}so, Valpara\'{i}so, Chile}


\begin{abstract}
While all models for the evolution of galaxies require the accretion of gas to sustain their growth via on-going star formation, it has proven difficult to directly detect this inflowing material. In this paper we use data of nearby star-forming galaxies in the SDSS IV Mapping Nearby Galaxies at Apache Point Observatory (MaNGA) survey to search for evidence of accretion imprinted in the chemical composition of the interstellar medium. We measure both the O/H and N/O abundance ratios in regions previously identified as having anomalously low values of O/H. We show that the unusual locations of these regions in the N/O vs.\ O/H plane indicate that they have been created through the mixing of disk gas having higher metallicity with accreted gas having lower metallicity. Taken together with previous analysis on these anomalously low-metallicity regions, these results imply that accretion of metal-poor gas can probably sustain star formation in present-day late-type galaxies.

\end{abstract}

\keywords{galaxies: abundances --- galaxies: evolution --- galaxies: statistics --- surveys --- techniques: imaging spectroscopy}

\section{Introduction} \label{sec:intro}

\subsection{Motivation}

Our understanding of the evolution of galaxies is that they continually grow with time. Part of this growth is through mergers between galaxies, but the majority of growth is through a more steady accretion of gas from the cosmic web \citep[e.g.,][]{Somerville_2015}. In galaxies currently forming stars, much of this accreted gas ultimately fuels the continuing formation of stars.

Despite the critical role played by accretion, direct observational evidence of significant inflows of gas into galaxies remains scarce outside the Milky Way \citep{Putman_2017}. The so-called down-the-barrel observations of intense star-forming galaxies using interstellar absorption lines as probes overwhelmingly reveal gas being expelled from galaxies \citep{Veilleux_2020}, with only a small minority of galaxies showing inflows \citep{Rubin_2012, Martin_2012, Zheng_2017}. This suggests that either the inflowing gas is not in the warm ionized phase being probed in absorption, or that it covers only a small solid angle as seen from the galaxy.   

In this paper we report the results of a complementary approach that provide evidence for on-going accretion of gas based on its chemical abundances. This study is a follow-up to our previous analysis of MaNGA data \citep{Hwang_2019} in which we found that late-type galaxies in the low-$z$ universe commonly have regions within them where the oxygen abundance (12 + log(O/H)) in the interstellar medium (ISM) is significantly lower than expected (relative to the tight empirical correlation between the oxygen abundance and the local stellar surface mass density for a fixed stellar mass). We defined these as anomalously low-metallicity (hereafter ALM, see section \ref{subsec:2.2}) regions, and showed that they correspond to regions of higher than average rates of star formation. We speculated that these ALM regions are sites in which low-metallicity gas has been recently accreted, gotten mixed with the pre-existing more metal-rich gas, and triggered star formation. 

The goal of this paper is to test this idea by measuring both O/H and N/O in the ALM regions to search for the tell-tale chemical signature of such an event. 

\subsection{Our Approach}


The key to our approach is the behavior of the ratio of nitrogen to oxygen (N/O) as a function of oxygen to hydrogen (O/H). Elements produced by nucleosynthesis are defined as primary or secondary. Primary elements are those whose yield is independent of the initial chemical composition of the star, while the yield of secondary elements depends on the initial composition of the star. Oxygen is a prototypical primary element, produced by Helium burning in massive stars ($> 10M_{\odot}$) and released promptly back to the ISM by core-collapse supernovae. Nitrogen production is more complex \citep[e.g.,][]{Arnett_1996, Kobayashi_2020}. It is produced from carbon and oxygen as part of the CNO cycle, primarily in intermediate-mass stars ($\sim 4 - 8 M_{\odot}$), and ejected into the ISM during the post-main-sequence asymptotic giant branch (AGB) phase. In a low metallicity star, nitrogen is produced from carbon and oxygen that were previously created via Helium burning in that star. In this case nitrogen behaves like a primary element, resulting in a fixed ratio of N/O. However, in higher metallicity stars, most of the carbon and oxygen used to create nitrogen were pre-existing in the ISM from which the star formed. In this case, nitrogen behaves like a secondary element, with N/O increasing with O/H \citep[e.g.,][]{Arnett_1996,Henry_2000,Meynet_2002,Matteucci_1985}. This pattern can be seen in data of chemical abundances of ionized gas in local star-forming galaxies \citep{Garnett_1990,Vila-Costas_1992,Henry_2000,Perez-Montero_2013,Andrews_2013,Vincenzo_2016,Belfiore_2017}. \citet{Andrews_2013} derived a schematic form for the N/O vs.\ O/H using SDSS star-forming galaxies (their Fig.~14): for 12 + log(O/H) less than 8.5, log(N/O) is constant $\sim -$1.4; while at higher values of 12 + log(O/H), they found N/O $\propto$ O/H$^{1.7}$.

This locus of N/O vs.\ O/H therefore traces the typical chemical evolution of the interstellar gas polluted by the by-products of stellar nucleosynthesis. In a simple closed-box model, a star-forming region in a galaxy evolves along this locus as more and more gas is converted into stars and the ISM metallicity rises. Eventually, the ISM metallicity becomes high enough for the significant production of secondary nitrogen. There are a number of ways in which a region in a galaxy could depart from this relation between O/H and N/O. For instance, in regions with a significant population of Wolf-Rayet stars, the gas can be polluted by nitrogen-rich winds from these stars \citep{Walsh_1989,Lopez-Sanchez_2007,Brinchmann_2008}. Fountain-flow models \citep{Bregman_1980, Fraternali_2008} in which outflowing metal-rich gas from the central region of a galaxy is accreted and mixed with metal-poor gas in the outer region can also produce gas that departs from the normal N/O vs.\ O/H relation \citep{Belfiore_2015}.

Of particular interest to the nature of the ALM regions, models in which infalling metal-poor gas is mixed with ambient metal-rich gas \citep{Koppen_2005} predict that the mixed gas will be characterized by having an unusually large N/O ratio for a given O/H value. Our specific goal in this paper is therefore to measure the locations of the ALM regions in the N/O vs.\ O/H plane. If, as suggested in \citet{Hwang_2019}, ALM regions represent sites where accretion of metal-poor gas has occurred, we should see an N/O excess. This technique has recently been used by \citet{AOH_2020} to show that accretion of metal-poor gas is likely to be responsible for triggering intense star formation in low-$z$ analogs to Lyman break galaxies.

In Section \ref{sec:data} we briefly describe the data used in our study, and we describe our approach to calculate the element abundances in Section \ref{sec:analysis}. In Section \ref{sec:results} we present our results and briefly discuss their implications in Section \ref{sec:implications}. We summarize our conclusions in Section \ref{sec:conclusions}.

\section{Data} \label{sec:data}

This work uses data from the MaNGA survey \citep{Bundy_2015,Yan_2016}, which is part of Sloan Digital Sky Survey (SDSS)-IV \citep{Blanton_2017,Gunn_2006}. MaNGA provides wavelength coverage from 3600 to 10300 \AA \ with a spectral resolution of R $\sim$ 2000 \citep{Drory_2015,Smee_2013}. The spectroscopic data used in this study come from the MaNGA Product Launches 9 \citep[MPL-9,][]{Wake_2017,Yan_2016_2}, an internal collaboration release comprising 8000 unique galaxies. The raw observed data are calibrated and sky-subtracted by the Data Reduction Pipeline (DRP) v2.7.1 \citep{Law_2016}, providing processed data cubes with spaxel sizes of 0.5 arcsec, which corresponds to a physical resolution of 1-2 kpc on the sky given the small redshifts of the targets. Emission line measurements (e.g., Gaussian fluxes and line-of-sight velocities) are provided by the Data Analysis Pipeline (DAP) v2.4.1 \citep{Westfall_2019,Belfiore_2019}. We use values measured after the stellar continuum has been fitted and subtracted. Galaxy properties such as redshift, total stellar mass, and axis ratio from elliptical Petrosian fitting (\textit{b/a}) are drawn from the NASA-Sloan catalog \citep{Blanton_2011}. The local stellar surface mass density is measured by the PIPE3D pipeline v2.3.1 \citep{Sanchez_2016a,Sanchez_2016b}.

\subsection{Sample selection} \label{subsec:sample}

Following the same criteria as in \citet{Hwang_2019}, we focus on late-type galaxies (selected using \texttt{fracdeV} $<$ 0.7, where \texttt{fracdeV} is the fraction of fluxes contributed from the de Vaucouleurs profile), pure star-forming spaxels (selected based on the [N\rom{2}] BPT diagram and the \citet{Kauffmann_2003} demarcation line) with deprojected local stellar surface mass density $>10^7M_{\odot}$ kpc$^{-2}$, total stellar mass $>10^9M_{\odot}$, and axis ratio from elliptical Petrosian fitting (\textit{b/a}) $>$ 0.3 (to exclude edge-on galaxies). We also require SNR $>$ 10 for the emission lines used in BPT diagrams and metallicity calculations.

\subsection{Definition of Anomalously Low-Metallicity (ALM) Regions} \label{subsec:2.2}

\citet{Hwang_2019} used the strong line calibrator [O\rom{2}]3727,3729/[N\rom{2}]6584 (O2N2) from \citet{Dopita_2013} to estimate the observed (12 + log(O/H))$_{\rm{obs}}$ of each spaxel. They computed the expected metallicity for each spaxel (12 + log(O/H))$_{\rm{exp}}$ via the interpolation of the tight relationship between local metallicity and the local stellar surface mass density at a given stellar mass \citep[$\Sigma_*-Z$ relation,][]{Barrera-Ballesteros_2016}. They then used the metallicity deviation $\Delta$log(O/H) = (12 + log(O/H))$_{\rm{obs}}$ $-$ (12 + log(O/H))$_{\rm{exp}}$ to define the ALM spaxels. Here we follow the same approach, but with a different strong line metallicity calibrator: RS32 = [O\rom{3}]5007/H$\beta$ + [S\rom{2}]6717,6731/H$\alpha$ from \citet{Curti_2020}. The photoionization models for the O2N2 indicator assume a relation between the O/H and N/O abundance ratios \citep{Dopita_2016}. Since we will compare the O/H and N/O of selected ALM spaxels, we need a calibrator that does not involve nitrogen. The RS32 calibrator is a good choice: it has only a mild dependence on the ionization parameter, and is insensitive to dust extinction due to the proximity of the emission line wavelengths\footnote{Extinction corrections to the emission line fluxes are described in \S\ref{subsec:extinction}}. \citet{Curti_2020} calibrated the RS32 diagnostic from electron temperature ($T_e$) based measurements of oxygen abundance in individual galaxies and stacked spectra, and we use the polynomial fitting results in their Table 2 to calculate (12 + log(O/H))$_{\rm{obs}}$ here.

\begin{figure}[h]
        \centering \includegraphics[width=\columnwidth]{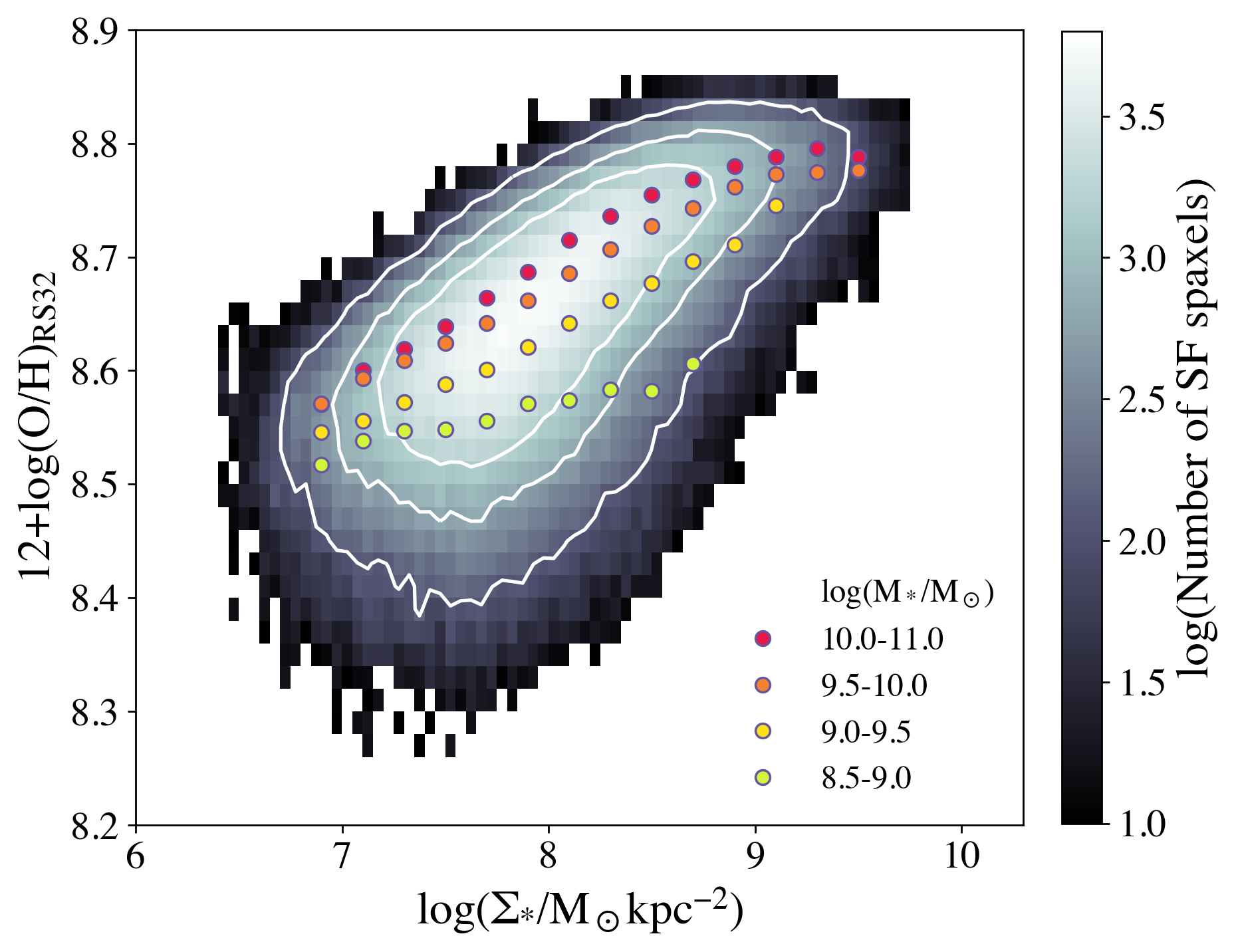}
        \caption{
                \label{fig:s-z} 
                Local relation between metallicity and stellar surface mass density. The circles represent the peaks of metallicity distributions in each stellar mass bin. The background distribution and contours show the distribution of all pure star-forming spaxels in late-type galaxies with stellar masses $>$ 10$^{8.5} M_{\odot}$. The bin of 10$^{8.5}$ $-$ 10$^{9} M_{\odot}$ is only used for interpolation, and we do not further analyze the galaxies in this bin.
        }
\end{figure}

\begin{figure}[h]
        \centering \includegraphics[width=\columnwidth]{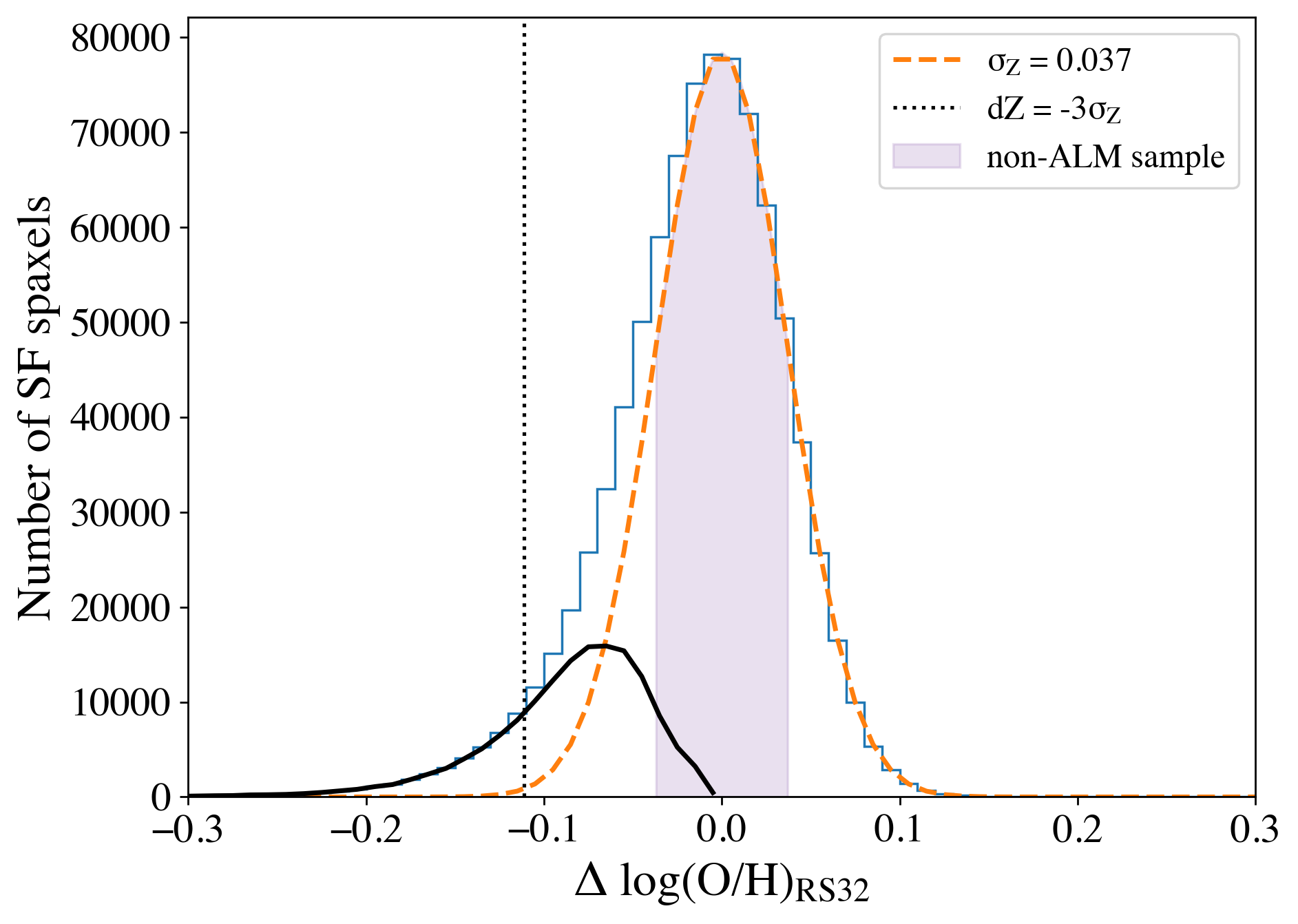}
        \caption{
                \label{fig:scatter} 
                Distribution of the metallicity deviation of star-forming spaxels in late-type galaxies with stellar mass $>10^9M_{\odot}$ (blue histogram). The orange dashed line is a Gaussian profile for “normal” spaxels, with a standard deviation $\sigma_Z = $ 0.037. The black solid line is the residual after subtracting the fitted Gaussian profile from the negative side of the histogram . The vertical dotted line marks where $\Delta \rm{log(O/H)} = -0.111$, and we refer to those spaxels deviating to lower metallicity by more than this value as ALM spaxels. The control sample (non-ALM) consisting of spaxels with $|\Delta \rm{log(O/H)}| \leqslant \sigma_Z$ are shaded in purple.
        }
\end{figure}

Our measured $\Sigma_*-Z$ relation is shown in Fig.\ref{fig:s-z}. Each circle represents the mode of the metallicity distribution inside each stellar mass and surface mass density bin. The step of the stellar mass bin is 0.5 dex, except for the bin of $10^{10}-10^{11}M_{\odot}$, because there are fewer galaxies with large stellar mass and because the mass-dependence of O/H is weak over this range \citep{Hwang_2019}. The bin of $M_*$ = 10$^{8.5}$ $-$ 10$^{9} M_{\odot}$ is only used for interpolation, and we do not further analyze the galaxies in this bin. The distribution of the metallicity deviation $\Delta$log(O/H) is shown as the blue histogram in Fig.~\ref{fig:scatter}. We observe the same low-metallicity tail as in \citet{Hwang_2019}. To characterize the ``normal" spaxels, we fit a Gaussian profile to the positive side of the metallicity deviation distribution, assuming a symmetric negative side. The best fit has $\sigma_Z=$ 0.037 and is shown as the orange dashed line in Fig.~\ref{fig:scatter}. The residual (black line in Fig.~\ref{fig:scatter}) is constructed by subtracting the fitted Gaussian from the negative side of the total distribution. The vertical dotted line marks where the metallicity deviation is three times the scatter. We define ALM spaxels as spaxels having $\Delta \rm{log(O/H)} < - 3\sigma_Z$ ($-$0.111 dex), which makes sure the ALM sample does not include substantial ``normal" spaxels characterized by the orange Gaussian profile. The control sample (non-ALM) then consists of spaxels where $|\Delta \rm{log(O/H)}| \leqslant \sigma_Z$ (shaded in purple in Fig.~\ref{fig:scatter}), meaning that they follow the local $\Sigma_*-Z$ relation well.

\section{Analysis} \label{sec:analysis}

\subsection{Extinction correction}\label{subsec:extinction}

Extinction corrections for emission line fluxes obtained from DAP assume the average Milky Way extinction curve in \citet{Fitzpatrick_2019} Table 3 (with $R(55)$ = 3.02, corresponding to the Milky Way mean value of $R(V)$ = 3.10), and the intrinsic flux ratio of H$\alpha$/H$\beta$ = 2.86 \citep[assuming $T_e = 10^4$K,][]{Osterbrock_2006}. We do not correct for extinction in the rare cases where the measured H$\alpha$/H$\beta <$ 2.86. Due to the typically small effect of the Milky Way reddening and the small redshifts of our sample galaxies, this extinction correction should be able to correct for extinction from both the Milky Way and the target galaxy. 

\subsection{Determination of O/H and N/O} \label{sec:3.2}

We analyzed both the individual spectrum of each spaxel and spectra created by stacking many spaxels in 0.1 dex bins of (12 + log(O/H))$_{\rm{obs}}$. By stacking the spectra we are able to increase the signal noise ratio (SNR) and thus detect the weak auroral lines required to apply the $T_e$-based direct method to measure O/H and N/O. However, the resulting systematic uncertainties prevent us from using the direct method metallicities to investigate the N/O vs.\ O/H relation (see Appendix A.5). Therefore, we continue to use the strong line diagnostics to determine element abundances\footnote{In spite of this, the stacked spectra enable us to search for faint and broad emission features produced by Wolf-Rayet stars in order to explore alternative sources of excess N/O (see \S4.2).}. The stacking procedure and direct method calculations are discussed in the Appendix.

For each spaxel, we first calculate the 12 + log(O/H) from the RS32 diagnostic as discussed in \S\ref{subsec:sample}. We then calculate N/O using the empirical relation found in \citet{AOH_2020}: 
\begin{equation}
\label{eq:no}
log(N/O) = 0.73 \times \rm{N2O2} - 0.58,
\end{equation}
where N2O2 = [N\rom{2}]6584/[O\rom{2}]3727, 3729 (corrected for extinction). Our estimations of O/H and N/O are quite independent as different lines are used. This helps to minimize any potential bias in our comparison of O/H and N/O. Note we are assuming here that the N/O ratio is the same as N$^+$/O$^+$. This is a safe assumption since the ionization potentials for O$^+$ and O$^{++}$ are very similar to those of N$^+$ and N$^{++}$ \citep{Garnett_1990}.

\subsection{Effects of metallicity calibrators}
There exist many strong line metallicity calibrators in the literature, and for the oxygen abundance ratio O/H, it is known that different metallicity calibrators can give results that differ by $\sim$0.4 dex \citep{Kewley_2008}. The nitrogen abundance ratio N/O is generally derived from the strong line ratio N2O2. The N/O relation used in the above section is derived based on Lyman break analogs, which are starburst galaxies with high specific star formation rate (sSFR), and our ALM spaxels also have high sSFR \citep{Hwang_2019}. Moreover, this N/O calibrator produces very similar results to other N/O relations in recent papers \citep[e.g.,][]{Strom_2017}.

While we explain why we choose to use RS32 in section \ref{subsec:2.2}, we here emphasize that our main result as shown in Fig. \ref{fig:nooh} below still hold with combinations of different O/H calibrators and N/O calibrators (see the test figure in Appendix B). The key point here is to be consistent with the calibrator used throughout (from the $\Sigma_*-Z$ relation calculation and sample selection to exploring the N/O vs.\ O/H relation), since our analysis depends on the \textit{relative} values of metallicities. We stick to our choice of RS32 for O/H because the R23 and O3S2 calibrators depends strongly on the ionization parameter, and they also show greater scatter around the fitted relation in the range of metallicity where most of our star-forming spaxels lie \citep[Fig. 1 in][]{Curti_2020}. Our main results are thus robust under different choices of the metallicity calibrators.

\begin{table*}[t]
\caption{\textbf{Left:} The initial and final metallicities after 60\% original gas + 40\% metal-poor gas (log(N/O) = $-$1.43 and half the initial O/H ratio) mixing (arrows in Fig.~\ref{fig:nooh}). \textbf{Right:} The initial and ending metallicities for the mixing model test described at the end of \S\ref{subsec:nooh} (arrows in Fig.~\ref{fig:exp}).}
\label{tab:mix}
\def\arraystretch{1.2} 
\centering
\begin{tabular}{c|c|c|c|c|c|c|c}
\hline
\multicolumn{4}{c|}{40\% accreted gas mixing} & \multicolumn{4}{c}{mixing model test}\\
\hline
\multicolumn{2}{c|}{before mixing} & \multicolumn{2}{c|}{after mixing} & \multicolumn{2}{c|}{initial point} & \multicolumn{2}{c}{ending point} \\
\hline
log(N/O) & 12 + log(O/H) & log(N/O) & 12 + log(O/H) & log(N/O) & 12 + log(O/H) & log(N/O) & 12 + log(O/H) \\
\hline
-0.742 & 8.75 & -0.839 & 8.65 & -0.742 & 8.75 & -1.04 & 8.60  \\
-1.04 & 8.65 & -1.11 & 8.55   & -1.04 & 8.65 & -1.24 & 8.50  \\
-1.34 & 8.55 & -1.36 & 8.45  & -1.34 & 8.55 & -1.42 & 8.39 \\
\hline
\end{tabular}
\end{table*}

\section{Results} \label{sec:results}

\subsection{The N/O vs.\ O/H plots}\label{subsec:nooh}

\begin{figure}[h]
        \centering \includegraphics[width=\columnwidth]{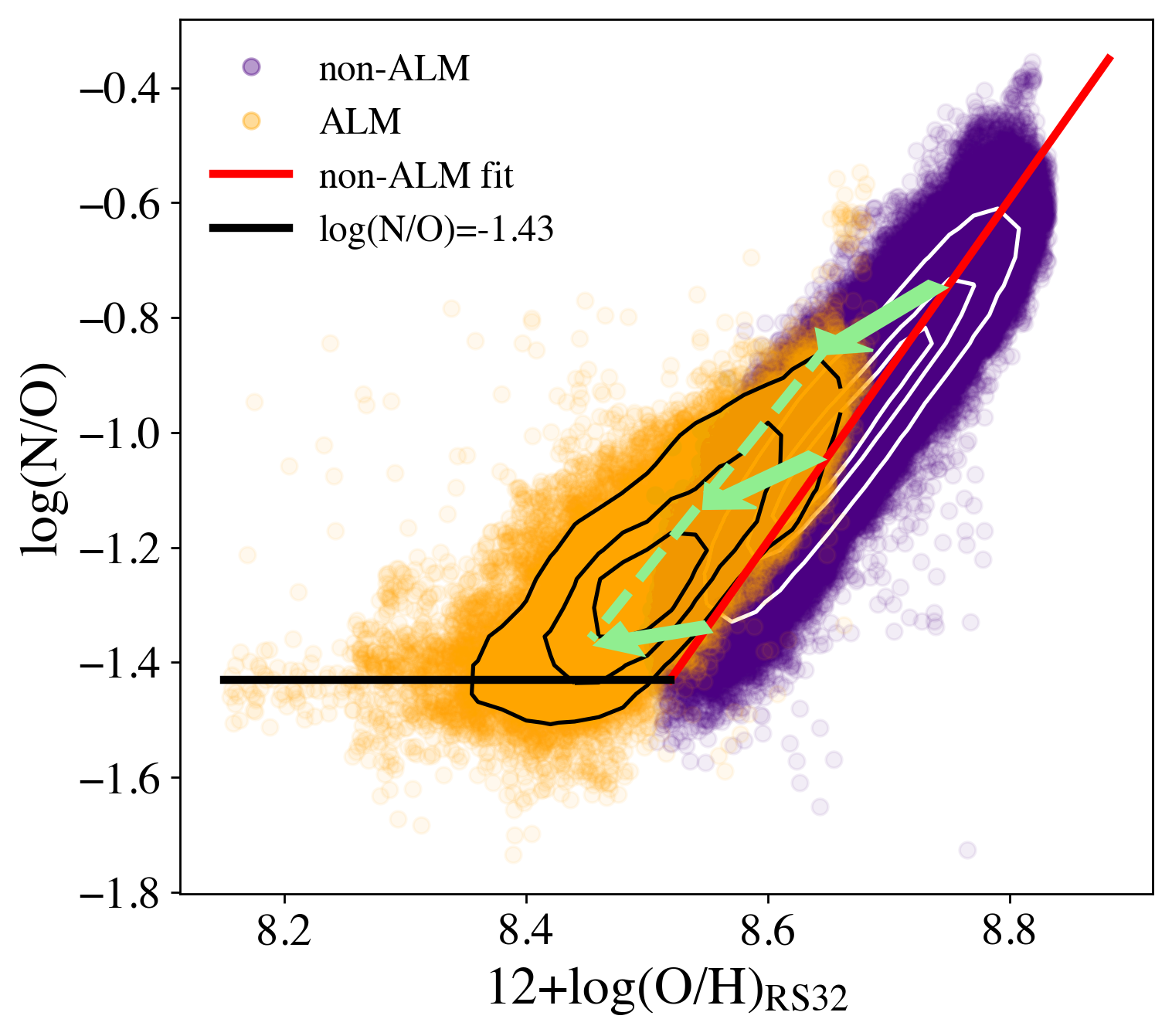}
        \caption{
                \label{fig:nooh} 
                ALM and non-ALM spaxels plotted on the N/O vs.\ O/H plane. The contours enclose 90\%, 60\%, 30\% percent of total data points, from the outermost line to the innermost line, respectively. Arrows show initial metallicities (ends of the arrows) and final metallicities (heads of the arrows) after 60\% original gas + 40\% low-metallicity gas (half the initial O/H ratio, log(N/O) = $-$1.43) mixing. We observe a clear offset between the distributions of ALM and non-ALM spaxels, and the hypothetical mixing model provides a good description to our results. 
        }
\end{figure}

We plot in Fig.~\ref{fig:nooh} the measured N/O vs.\ O/H for the ALM ($\Delta \rm{log(O/H)} < - 3\sigma_Z$) and non-ALM ($|\Delta \rm{log(O/H)}| \leqslant \sigma_Z$) spaxels. The non-ALM spaxels clearly show the trend of N/O increasing with O/H, and the best-fit line is 
\begin{equation}
\label{eq:2}
    log(N/O)= 2.98 \times [12+log(O/H)] - 26.82.
\end{equation}
The ALM spaxels partially overlap with the non-ALM spaxels, and show a plateau of N/O $\sim$ $-$1.4 at low O/H ($\lesssim$ 8.5) as expected. We thus take the log(N/O) value from \citet{Andrews_2013} for 12 + log(O/H) $\leq$ 8.5:
\begin{equation}
\label{eq:1}
    log(N/O)=-1.43.
\end{equation}

As can be seen, there is a clear offset between the loci of the ALM and non-ALM spaxels. The ALM spaxels lie systematically off the normal relationship, having abnormally large N/O for a given O/H. We interpret this as a signature of the mixing of metal-rich gas (on the linear part of the N/O vs.\ O/H relation) with metal-poor gas (on the flat part of the N/O vs.\ O/H relation). To further quantify this, we calculate log(N/O) and 12 + log(O/H) for a hypothetical gas mixing scenario: We assume that pre-existing gas in the disk with different initial metallicities (following Eq.~\ref{eq:2}) mixes with accreted low-metallicity gas with half the initial O/H ratio and log(N/O) = $-$1.43 (following Eq.~\ref{eq:1}). These abundances in the accreted gas are consistent with measurements of the circumgalactic medium \citep{Prochaska_2017}, which represents the likely gas source for accretion. We consider the case where the accreted gas mass is 40\% of the total gas mass. The resulting metallicities of mixing are summarized in Table~\ref{tab:mix} and shown as green arrows in Fig.~\ref{fig:nooh}. Note that this is meant to be a simple illustrative example of how this mixing process works, and different choices could be made for the chemical composition of the accreted gas.

\begin{figure}
        \centering \includegraphics[width=\columnwidth]{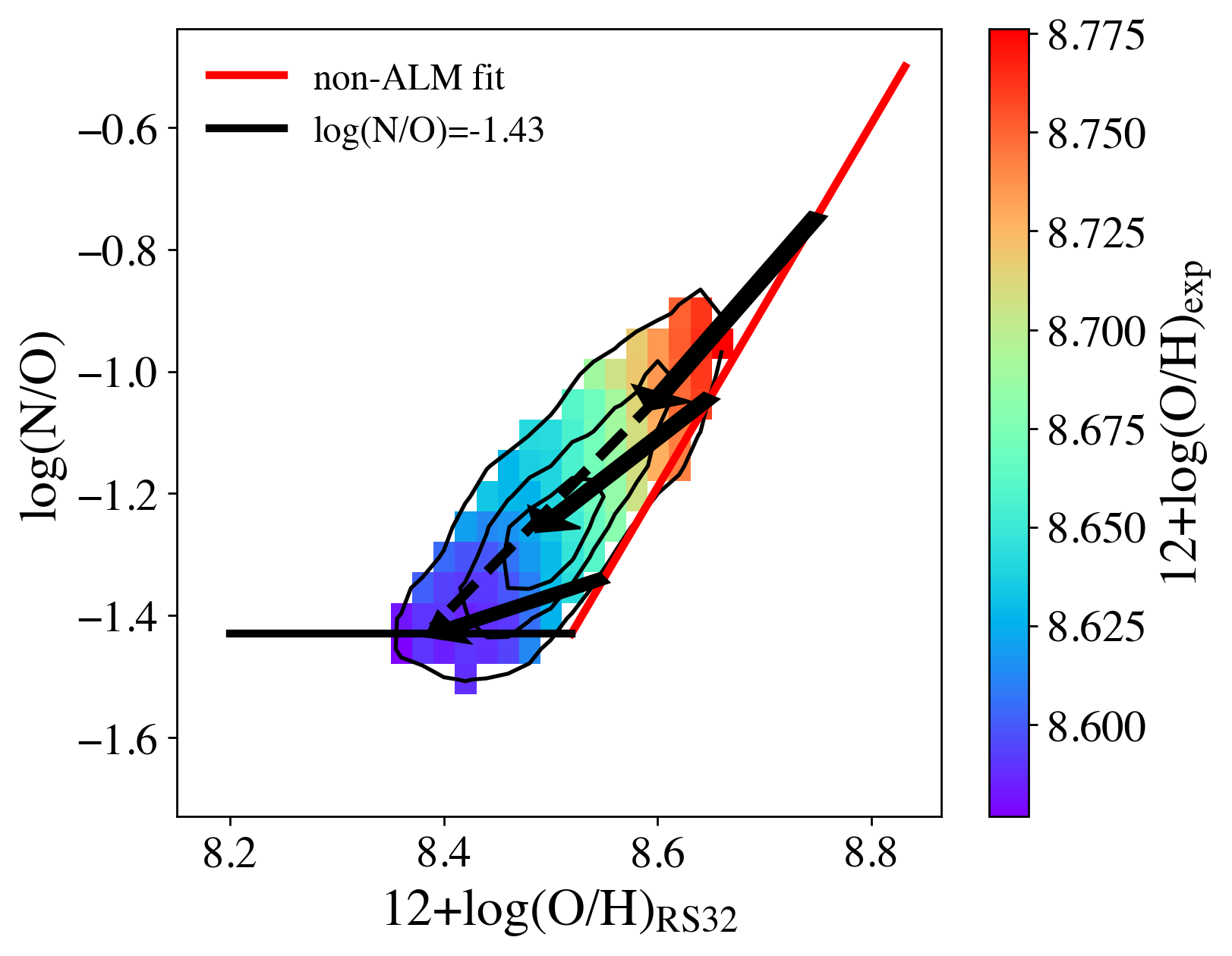}
        \caption{
                \label{fig:exp} 
                ALM spaxels color-coded by (12 + log(O/H))$_{\rm{exp}}$, their expected initial oxygen abundance based on the local $\Sigma_*-Z$ relation (Fig.~\ref{fig:s-z}). The contours enclose 90\%, 60\%, 30\% percent of total data points, from the outermost line to the innermost line, respectively. The arrows connect initial points which lie along the normal distribution line to the centers of the ALM spaxel distributions with the same (12 + log(O/H))$_{\rm{exp}}$ as the initial points. The results are qualitatively similar to the simple mixing model shown in Fig.~\ref{fig:nooh}. See text for details.
        }
\end{figure}

This mixing model can be further tested by using (12 + log(O/H))$_{\rm{exp}}$ values calculated in \S\ref{subsec:2.2} to connect the origins of the gas prior to accretion (lying along the non-ALM fit line) to its present location in the N/O vs.\ O/H plot. That is, we assume that the initial O/H value of an ALM spaxel prior to mixing is the value expected based on the $\Sigma_*-Z$ relation (Fig.~\ref{fig:s-z}) and that the initial value of N/O is given by Eq.~\ref{eq:2}. For each such initial O/H value, we locate all the ALM spaxels whose (12 + log(O/H))$_{\rm{exp}}$ is within 0.02 dex of this initial value, and take the center of their distribution in the N/O vs.\ O/H plot as the ending point of the mixing process. The resulting values for O/H and N/O are summarized in Table~\ref{tab:mix}. In Fig.~\ref{fig:exp}, we plot ALM spaxels within the contour which encloses 90\% of the total data points, color-coded by (12 + log(O/H))$_{\rm{exp}}$, and show arrows connecting the initial points (same as those in Fig.~\ref{fig:nooh}) to the ending points. We see that the trajectories of the ALM spaxels are in qualitative agreement with the simple mixing model in Fig.~\ref{fig:nooh}. This provides further support for the interpretation of the ALM spaxels as regions where pre-existing higher-metallicity gas mixes with accreted lower-metallicity gas. 

\begin{figure}
        \centering \includegraphics[width=\columnwidth]{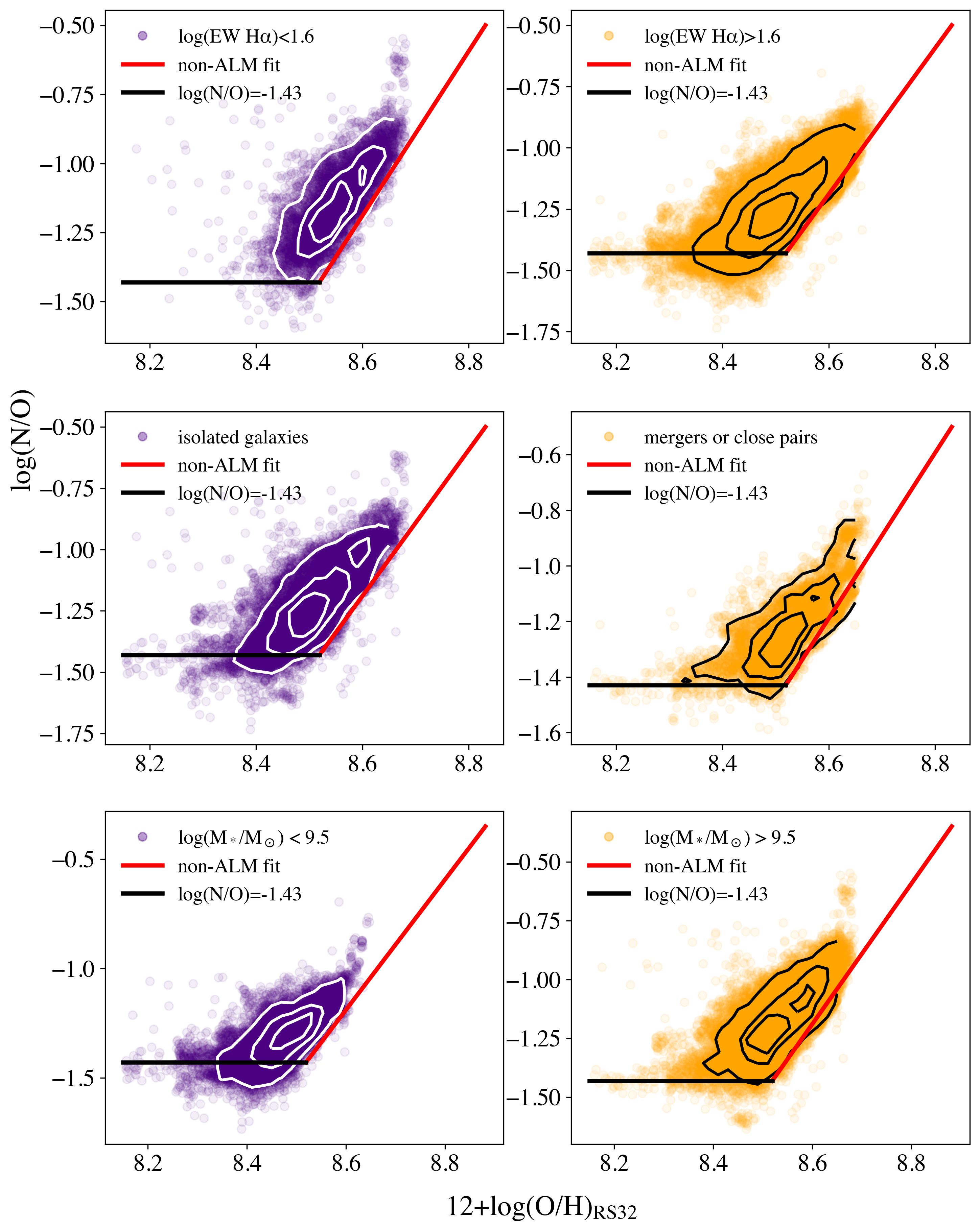}
        \caption{
                \label{fig:sub} 
                ALM spaxels subdivided based on the equivalent width of the H$\alpha$ emission line (top), whether the galaxy was classified as merger/close pairs or isolated (middle), and the stellar mass of the galaxy (bottom). The contours in each panel enclose 90\%, 60\%, 30\% percent of total data points plotted in that panel, from the outermost line to the innermost line, respectively. We observe the N/O excess characteristic of gas mixing in all cases. 
        }
\end{figure}

To test the robustness of these results, we subdivide the ALM sample into three different pairs of smaller samples, based on: 1) the equivalent width of the H$\alpha$ emission line, 2) the stellar mass of the galaxy, and 3) whether the galaxy was classified as merger/close pairs or isolated in \citet{Hwang_2019}, where it was found that the incidence rate of ALM spaxels is correlated with all these properties. We plot the sub-samples for comparison in Fig.~\ref{fig:sub}. Each sub-sample shows enhanced N/O at fixed O/H compared to the expected distribution with no clear differences between the sub-samples of each analyzed pair.

\begin{figure}
    \centering
    \includegraphics[width=\columnwidth]{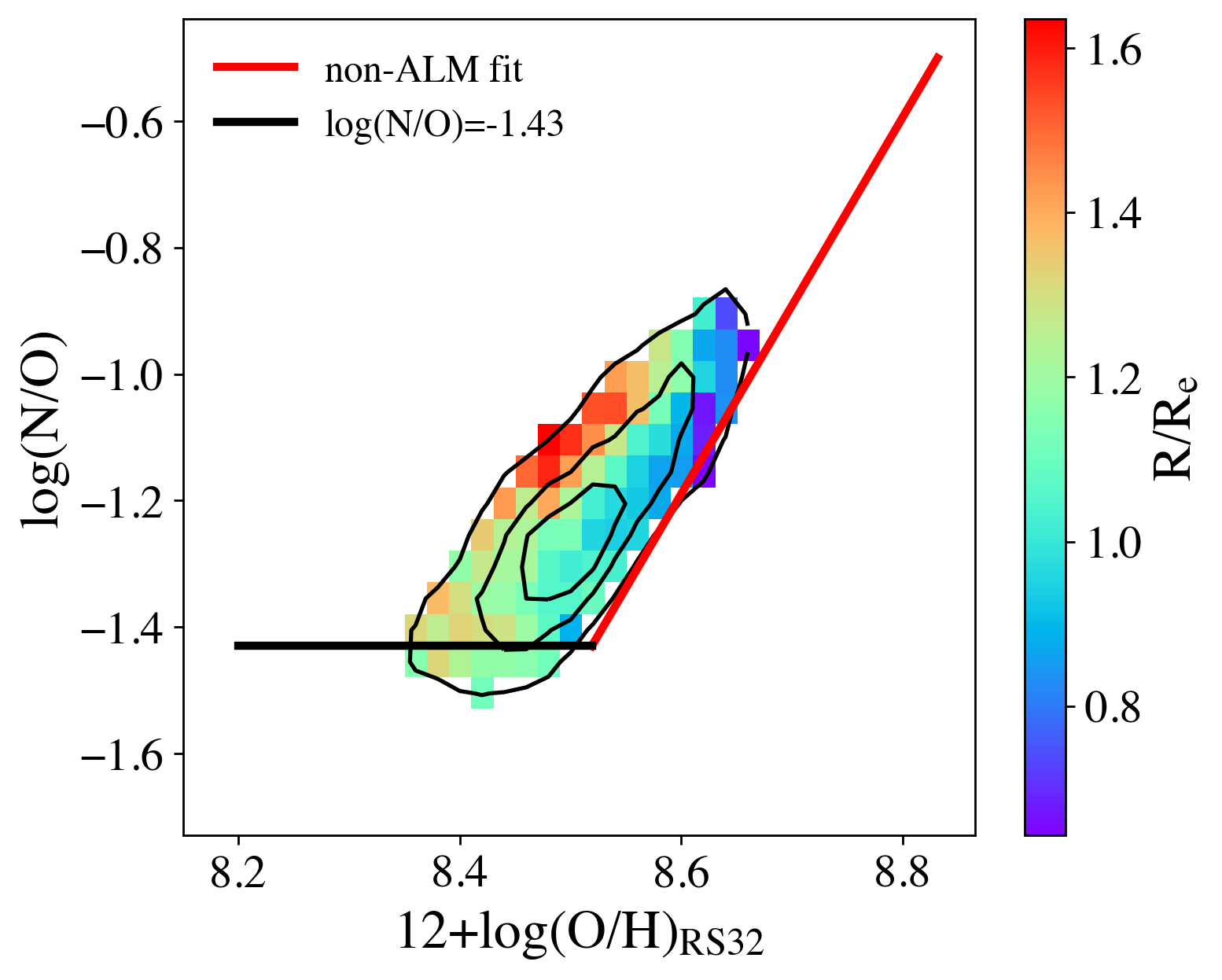}
    \caption{
    ALM spaxels color-coded by $R/R_e$, the distance to the center of the galaxy normalized by the galaxy half-light radius. The contours enclose 90\%, 60\%, 30\% percent of total data points, from the outermost line to the innermost line, respectively. Spaxels with the largest offset from the normal relation are preferentially found in the outer region ($R > 1.5 R_e$).
    }
    \label{fig:nooh_re}
\end{figure}

One particularly interesting property of the ALM spaxels is their location in the galaxy. \citet{Hwang_2019} showed that they occur preferentially at larger radii than the normal spaxels. In Fig.~\ref{fig:nooh_re} we plot all ALM spaxels in the N/O vs.\ O/H plane color-coded by the radius at which the spaxel is located, normalized by the galaxy half-light radius ($R/R_e$). We see that the ALM spaxels which deviate most strongly from the normal N/O vs.\ O/H  relation (e.g., the points lying along the left edge of the distribution in the figure) show a strong trend to be located in the outer regions of their host galaxies ($R > 1.5 R_e$). In the context of the mixing model, these locations in the N/O vs.\ O/H plane would correspond to a a more dominant fraction of accreted vs.\ pre-existing gas in these locations. This is qualitatively consistent with a model where gas is accreted from outside the galaxy.

\begin{figure*}
        \centering \includegraphics[width=2\columnwidth]{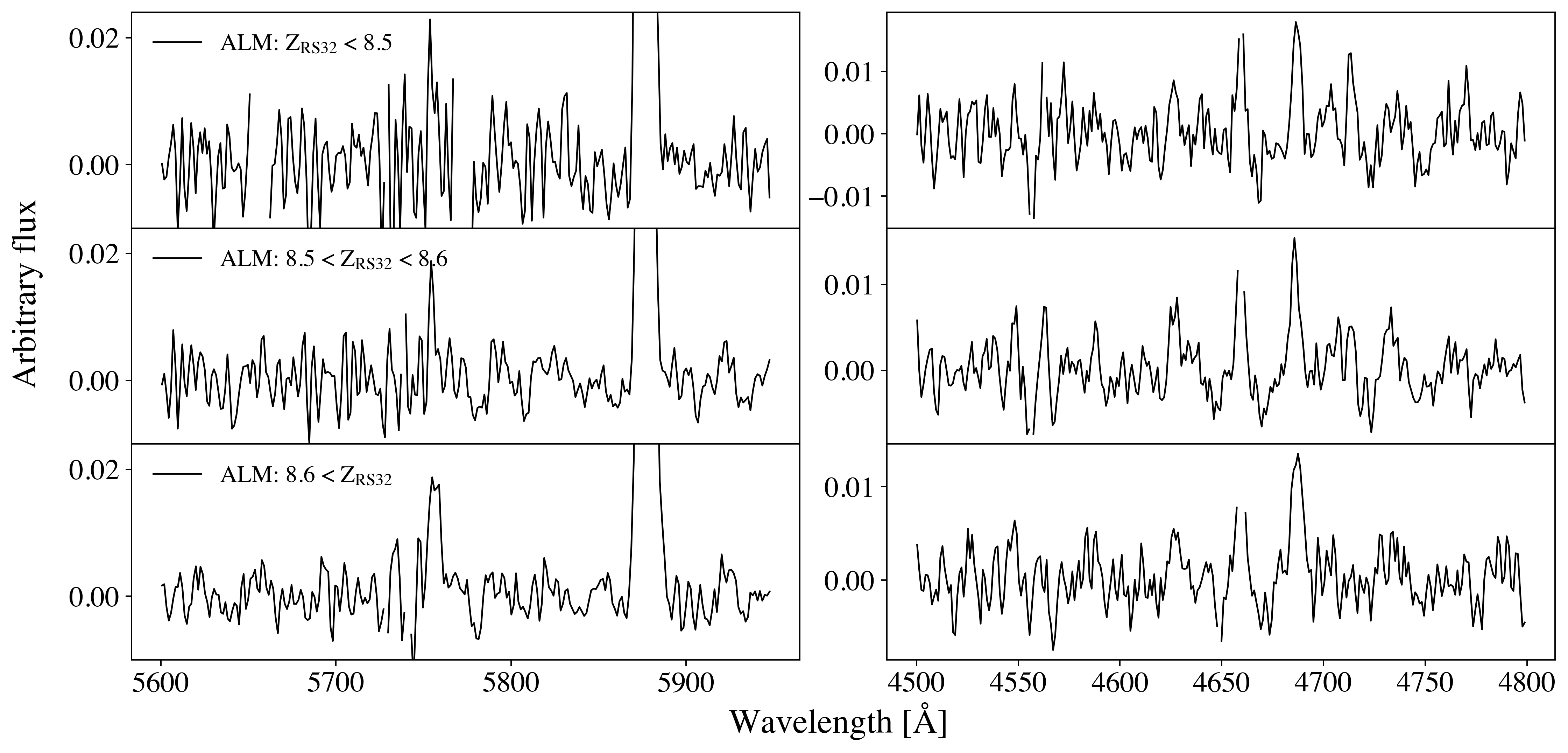}
        \caption{
                \label{fig:wr} 
                Stellar continuum subtracted ALM stacked spectra in regions of WR features. We do not detect the WR feature of broad emission bumps.
        }
\end{figure*}

\subsection{Other possible causes of N/O excess}\label{subsec:4.2}

The fact that the ALM spaxels have an excess N/O for a given O/H rules out the interpretation that they are simply less chemically-evolved regions (in which case they would have lower O/H but still follow the normal N/O vs.\ O/H relation). However, as noted in the introduction, there are alternative ways in which gas in a galaxy could depart from the normal N/O vs.\ O/H relation besides the accretion of metal-poor gas. We briefly discuss the possibilities of these alternatives here.

First, strong winds carrying nitrogen-enriched gas from Wolf-Rayet (WR, post-main sequence hot and massive) stars can raise N/O in the ISM \citep{Walsh_1989,Lopez-Sanchez_2007}. We use the high SNR stacked spectra described in the Appendix to search for the two strongest features among the series of weak and broad emission features due to WR stars, at $\sim$ 4660 \AA\ and $\sim$ 5810 \AA. In Fig.~\ref{fig:wr} we show zoomed-in views of these regions in stellar continuum subtracted stacked spectra of ALM spaxels. In no case do we detect the WR feature of broad emission bumps (for examples of such broad emission bumps, see Fig.~20 in \citet{AOH_2020}). Note that \citet{Brinchmann_2008} found an enhancement in N/O by $\sim$ 0.1 dex in galaxies with detectable WR features compared to those without. This suggests that the ISM can be enriched in nitrogen during a phase in which a strong WR population is present. Since we do not detect a strong WR population in the ALM regions, there is no evidence that WR stars are responsible for the enhanced N/O we see.

Second, the release of secondary nitrogen by post-main-sequence intermediate mass stars \citep[4$-$7$M_{\odot}$,][]{Kobayashi_2020} lags the release of oxygen in more massive stars, since these intermediate mass stars have lifetimes of $\sim 10^8$ years. An excess in N/O could be produced in the case of a strong post-starburst system in which no core-collapse supernovae are releasing newly created oxygen, but the intermediate mass stars produced 10$^8$ years ago in the burst are finally returning secondary nitrogen \citep[e.g.,][]{Schaefer_2020}. A selective loss of oxygen via supernovae-driven galactic-winds during the burst phase could also contribute to a high N/O value during the post-burst phase \citep{Vincenzo_2016}. However, we emphasize that this scenario is inconsistent with the properties of the ALM spaxels, which are shown to be tracing on-going bursts of star formation with a significant population of short-lived massive stars \citep{Hwang_2019}.

Finally, an interesting variant of the mixing scenario is one in which metal-rich gas is ejected from the central region of a galaxy as a fountain-flow and mixes with relatively metal-poor gas in the outer disk \citep{Belfiore_2015}. While this can produce gas with enhanced N/O, it would also enhance O/H and lead to regions with anomalously high metallicity. 

Therefore, we conclude that the accretion of metal-poor gas is responsible for the chemical composition of gas in the ALM regions in our case. 

\section{Implications} \label{sec:implications}

Our results strongly support the interpretation proposed by \citet{Hwang_2019} in which the anomalously low-metallicity (ALM) regions are the sites at which low-metallicity accreted gas has been mixed into the pre-existing higher-metallicity interstellar medium. Such infall of pristine gas has also been recognized as an explanation for the fundamental metallicity relation (FMR), which states that metallicity decreases sharply with increasing star formation rate at low stellar mass \citep{Mannucci_2010}. Similar conclusions were also reached by \citet{Sanchez-Almeida_2018} and \citet{Sanchez-Menguiano_2019} who found a local anti-correlation between gas metallicity and star formation rate in a large population of local star-forming galaxies.

With this in mind, we briefly summarize the implications this would have for the evolution of star-forming galaxies in the current epoch. \citet{Hwang_2019} showed that the incidence rate of ALM regions in star-forming galaxies increased strongly with both the galaxy-wide and local (ALM regions) values of the specific star formation rate (SFR/$M_*$). Given the evidence we have presented here, this strongly suggests that this enhancement in star formation is causally related to the accretion of low-metallicity gas. This interpretation is also consistent with their results that the ALM incidence rate was higher in the outer regions of galaxies, and in mergers and close pairs of galaxies. The former is expected for gas accreted from outside the galaxy, while the latter suggests that in some cases the accretion event was associated with mass transfer from a smaller nearby companion. Roughly 75\% of galaxies with ALM regions in \citet{Hwang_2019} were not mergers or in close pairs, leading them to argue that in most galaxies ALM regions are associated with accretion of gas from the circumgalactic medium rather than from mass transfer.

Under the assumption that ALM regions represent accretion, \citet{Hwang_2019} roughly estimated what the implied accretion rates would be. Based on the estimated lifetimes of the ALM regions of a few hundred Myr, and taking the incidence rate of ALM regions to represent the duty cycle of such events, they showed that the time-averaged accretion rate of ALM gas was similar to the star formation rate in star forming galaxies over the mass-range $M \sim 10^9$ to $10^{10} M_{\odot}$.

With our additional evidence supporting the identification of ALM regions as the sites of accretion, we believe that there is now strong (though still indirect) evidence that star formation in star-forming galaxies in the present epoch can be sustained and triggered by the delivery of gas into the galaxies. These results help to validate one of the fundamental components in models of galaxy evolution. 

\section{Conclusions} \label{sec:conclusions}

There have been powerful arguments that the stellar components of galaxies are built through the accretion of gas, initially residing in the cosmic web. The influx of fresh gas is believed to regulate the rate at which this gas is consumed via star formation or expelled by galactic winds. However, direct kinematic evidence for infalling gas into galaxies (e.g., red-shifted interstellar absorption lines) has been scarce. In this paper we tested the idea that the chemical composition of gas in galaxies can provide strong indirect evidence for gas accretion.

\cite{Hwang_2019} used data from the SDSS IV MaNGA surevy to document the wide-spread occurrence of regions with anomalously low-metallicity (ALM) in the disks of star-forming low-$z$ galaxies. They proposed that these regions correspond to the sites of localized and sporadic accretion events in which low-metallicity gas is mixed with the pre-existing high-metallicity ISM. In this paper we test these predictions by measuring and comparing N/O and O/H for the ALM regions. 

\begin{enumerate}
\item We utilize integral field spectroscopy (IFS) data in the SDSS IV MaNGA MPL-9 data release, and estimated the oxygen abundance 12 + log(O/H) using the RS32 strong line method in \citet{Curti_2020}. We estimate log(N/O) following the relationship found with local Lyman break analog galaxies in \citet{AOH_2020}.

\item Based on the tight empirical relation between metallicity and local stellar surface mass density at fixed stellar mass, we identified ALM regions as spaxels with observed metallicity lower than the expected value by more than 0.111 dex. This resulted in 37288 ALM spaxels in 685 star-forming galaxies with $M_* > 10^9 M_{\odot}$. 

\item Our results show that the ALM regions lie in the region of the N/O vs.\ O/H plot where N/O is unusually large for a given O/H value. This scenario is in agreement with a mixing (accretion) model and thus rules out the alternative interpretation that the ALM regions are just regions that are less chemically-evolved due to less prior star formation resulting in less metal-enrichment (in which case they would follow the same N/O vs.\ O/H relation as the normal non-ALM spaxels).

\item We consider other possible ways that could lead to enhanced N/O at given O/H, such as Wolf-Rayet stars, post-starburst systems, and fountain flows. We find them incompatible with other properties of our ALM regions and we therefore conclude that the accretion of metal-poor gas is responsible for the chemical composition of gas in the ALM regions.
\end{enumerate}

\citet{Hwang_2019} showed that ALM regions are associated with regions of unusually high star formation rates, and suggested that star formation in late-type galaxies at low-$z$ can be sustained and even stimulated by the accretion events they trace. Our results provide confirmation that ALM regions are indeed accretion sites, and thereby supported the idea that we are witnessing galaxy building in action.

R.R thanks National Council for Scientific and Technological Development (CNPq), Coordenação de Aperfeiçoamento de Pessoal de N\'{i}vel Superior (CAPES) and Fundação de Amparo à Pesquisa do Estado do Rio Grande do Sul (FAPERGS). J.G.F-T is supported by Fondo Nacional de Desarrollo Cient\'{i}fico y Tecnol\'{o}gico (FONDECYT) No. 3180210. R.M acknowledges ERC Advanced Grant 695671 "QUENCH" and support by the Science and Technology Facilities Council (STFC).

Funding for the Sloan Digital Sky Survey IV has been provided by the Alfred P. Sloan Foundation, the U.S. Department of Energy Office of Science, and the Participating Institutions. SDSS acknowledges support and resources from the Center for High-Performance Computing at the University of Utah. The SDSS web site is \url{www.sdss.org}.

SDSS is managed by the Astrophysical Research Consortium for the Participating Institutions of the SDSS Collaboration including the Brazilian Participation Group, the Carnegie Institution for Science, Carnegie Mellon University, Center for Astrophysics $|$ Harvard \& Smithsonian (CfA), the Chilean Participation Group, the French Participation Group, Instituto de Astrofísica de Canarias, The Johns Hopkins University, Kavli Institute for the Physics and Mathematics of the Universe (IPMU) / University of Tokyo, the Korean Participation Group, Lawrence Berkeley National Laboratory, Leibniz Institut für Astrophysik Potsdam (AIP), Max-Planck-Institut für Astronomie (MPIA Heidelberg), Max-Planck-Institut für Astrophysik (MPA Garching), Max-Planck-Institut für Extraterrestrische Physik (MPE), National Astronomical Observatories of China, New Mexico State University, New York University, University of Notre Dame, Observatório Nacional / MCTI, The Ohio State University, Pennsylvania State University, Shanghai Astronomical Observatory, United Kingdom Participation Group, Universidad Nacional Autónoma de México, University of Arizona, University of Colorado Boulder, University of Oxford, University of Portsmouth, University of Utah, University of Virginia, University of Washington, University of Wisconsin, Vanderbilt University, and Yale University. 

\software{Data Reduction Pipeline (DRP) v2.7.1 \citep{Law_2016}, Data Analysis Pipeline (DAP) v2.4.1 \citep{Westfall_2019,Belfiore_2019}, PIPE3D pipeline v2.3.1 \citep{Sanchez_2016a,Sanchez_2016b}, marvin \citep{marvin_2019}}


\appendix

\section{Direct method metallicity}
Here we present our attempt to apply the direct $T_e$-based method to calculate metallicies. We discuss our spectra stacking and calculation procedures, and explain why we do not end up using these results.\\

\subsection{Spectra stacking}
In order to measure the fluxes of weak auroral lines, we must stack spectra of ALM spaxels to obtain an adequate SNR. 
We select three metallicity bins for stacking defined using previously calculated (12 + log(O/H))$_{\rm{obs}}$ from the RS32 calibrator: Z$_{RS32}$ $<$ 8.5, 8.5 $<$ Z$_{RS32}$ $<$ 8.6, 8.6 $<$ Z$_{RS32}$. The spectra to be stacked are taken from unbinned DRP products accessed via \texttt{marvin} \citep{marvin_2019}. Before stacking, we check both \texttt{DRP Maskbits} and \texttt{DAP Maskbits} from \texttt{marvin} and exclude problematic spaxels with bad quality flags. We then follow \citet{Andrews_2013} to stack our spectra. 

We first shift each spectrum to the rest-frame using its redshift and the local line-of-sight emission line velocity obtained from DAP maps. We then interpolate each spectrum in a log-linear way with $\Delta$log$\lambda$ = 10$^{-4} \rm{\AA}$, between the minimal and maximum wavelength out of all spectra in the corresponding metallicity bin. Finally we normalize the spectra to the mean flux between 4400$-$4450 \rm{\AA} and co-add (average the flux at each wavelength) all the spectra in the metallicity bin. 

\subsection{Flux measurements}
To analyze the stacked spectra, we use MaNGA DAP software package v2.4.1\footnote{\url{https://github.com/sdss/mangadap}} \citep{Westfall_2019,Belfiore_2019}, which automatically fits and subtracts the stellar continuum, providing measured emission line fluxes and corresponding flux errors. We use the \texttt{MILES-HC} template for the stellar continuum fitting and the \texttt{MASTAR-HC} template for emission line fitting, following the same setting in MaNGA MPL-9 data analysis.

The features of interest are the weak auroral lines used to measure $T_e$: [O\rom{3}]4363, [N\rom{2}]5755, [O\rom{2}]7320, 7331, and the regions between 4600$-$4700 \AA\ and 5750$-$5850 \AA\ where the strongest Wolf-Rayet emission features are located. We use the package to fit the stellar continuum and emission lines for only 200 \rm{\AA} excerpts from the complete spectrum, centered on the auroral line(s) of interest. For other stronger emission lines, the results are from fitting the stellar continuum over the entire spectrum. Ideally we would expect SNR to increase by a factor of $\sqrt{N}$ when stacking N individual spectrum together, however, we find in our case that increasing the number of spectra used in stacking could barely increase SNR once the number exceeds $\sim$ 1000, i.e., there is irreducible systematic noise in the original spectra which we use for stacking, probably due to imperfect starlight subtraction. Nevertheless, the stacking procedure described above allows us to successfully measure the fluxes of the auroral lines. The extinction is corrected following \S\ref{subsec:extinction} above.

\subsection{Electron temperature and density}

In order to determine the electron temperature ($T_e$), we need to apply corrections based on the electron density. The electron density $n_e$ is estimated from the ratio of [S\rom{2}]6716, 6731 lines, using equation (5) in \citet{Yates_2020}. We calculate separate electron temperatures for O\rom{2} and O\rom{3} ($T_e$(O\rom{3}) and $T_e$(O\rom{2})) by solving equation (3) in \citet{Yates_2020} iteratively. Our calculation process agrees with Yates' statement that convergence is typically achieved within three iterations for $T_e$(O\rom{3}), and five iterations for $T_e$(O\rom{2}). The auroral line [N\rom{2}]5755 is too weak to measure even in the stacked spectra, so we assume $T_e$(N\rom{2}) = $T_e$(O\rom{2}). This assumption is supported by the comparisons done by \citet{Curti_2017}. In estimating the uncertainties of electron temperatures, we assume all uncertainties stem from the flux uncertainties of the auroral lines [O\rom{3}]4363, [O\rom{2}]7320, 7331 as other stronger emission lines have much smaller flux uncertainties. 

The uncertainties in the fluxes of the weak auroral lines are dominated by systematic errors in the subtraction of features in the stellar continuum. Therefore, rather than use the pipeline-estimated errors, we calculate the flux uncertainty in these lines as the standard deviation of the fluxes in the residual (starlight-subtracted) spectrum integrated over wavelength bins with size roughly equal to the FWHM of the corresponding line (10 \rm{\AA} for [O\rom{3}]4363, 20 \rm{\AA} for the blended [O\rom{2}]7320, 7331). This standard deviation was calculated around each line(s) over the the 200 \rm{\AA} fitting window. The uncertainty of the electron temperature is then calculated using a Monte Carlo method where the observed auroral line flux is deviated from its measured value by a Gaussian with $\sigma$ equal to its flux uncertainty, and an electron temperature is calculated with this new flux. This process is repeated 1000 times and the standard deviation of the resulting distribution is taken as the uncertainty of the measured electron temperature.


\subsection{Determination of O/H and N/O}

Direct measurements of singly- and doubly-ionized oxygen abundances O$^+$/H$^+$ and O$^{++}$/H$^+$ are calculated using equation (6) and (7) in \citet{Yates_2020}, respectively. The uncertainties in oxygen abundances are derived with the same Monte Carlo method described before, propagating the previously found uncertainties in the electron temperatures. The total oxygen abundance is obtained via O/H = O$^+$/H$^+$ + O$^{++}$/H$^+$. 

To obtain direct measurements of the N/O ratio, We assume log(N$^+$/O$^+$) = log(N/O) and calculate it following equation (3) and (6) in \citet{Izotov_2006}. 

\subsection{Systematic Uncertainties}

The uncertainties in the direct method metallicities estimated as discussed above is on the order of 0.01 dex. However, we find larger systematic uncertainties in the direct method (12 + log(O/H))$_{\rm{direct}}$. As we already know that increasing the number of raw spectra in the stacking could no longer improve the SNR of the stacked spectrum once the number of raw spectra exceeds $\sim$ 1000, we try to make 10 stacked spectra in metallicity bins Z$_{RS32}$ $<$ 8.5 (not many spectra fall here), 8.5 $<$ Z$_{RS32}$ $<$ 8.6, 8.6 $<$ Z$_{RS32}$ $<$ 8.7, each with $\sim$ 2000 raw spectra drawn randomly from all spaxels satisfying the criteria described in \S\ref{subsec:sample} in that metallicity bin. We calculate (12 + log(O/H))$_{\rm{direct}}$ for the 10 stacked spectra in each of the metallicity bin, and tabulate their mean and standard deviation in Table \ref{tab:std}.

\begin{table}[h]
\caption{Table summarizing the mean and standard deviation of (12 + log(O/H))$_{\rm{direct}}$ for the 10 stacked spectra in each of the three metallicity bins.}
\label{tab:std}
\def\arraystretch{1.3} 
\setlength\tabcolsep{12pt} 
\centering
\begin{tabular}{c P{1.6cm} P{1.4cm}}
\hline
metallicity bin & mean oxygen abundance & standard deviation  \\
 \hline
Z$_{RS32}$ $<$ 8.5 & 8.25 &  0.0145  \\
8.5 $<$ Z$_{RS32}$ $<$ 8.6 & 8.38 & 0.0942  \\
8.6 $<$ Z$_{RS32}$ $<$ 8.7 & 8.40 & 0.109  \\
\hline
\end{tabular}
\end{table}


We observe large scatter in the measured direct method oxygen abundances in the higher metallicity bins. While the scatter is significantly smaller in the Z$_{RS32}$ $<$ 8.5 bin, 12 + log(O/H) values in this range lie on the flat part of the N/O vs.\ O/H plot and are thus poor diagnostics of the mixing model. Moreover, the standard deviation of oxygen abundances calculated from the stacked spectra is about 0.1 dex in the two higher metallicity bins. By our definition of ALM spaxels in \S\ref{subsec:2.2}, the difference in 12 + log(O/H) values between ALM and non-ALM spaxels is only $\sim$0.111 dex. Given such large systematic uncertainties in the direct method oxygen abundance, we are unable to reliably distinguish ALM and non-ALM spaxels in the N/O vs.\ O/H plane. Although we therefore use the strong line diagnostics in our chemical composition analysis, the high SNR stacked spectra helped us to look for Wolf-Rayet star emission features as discussed in \S\ref{subsec:4.2}.

\begin{figure}[h]
        \centering \includegraphics[width=\columnwidth]{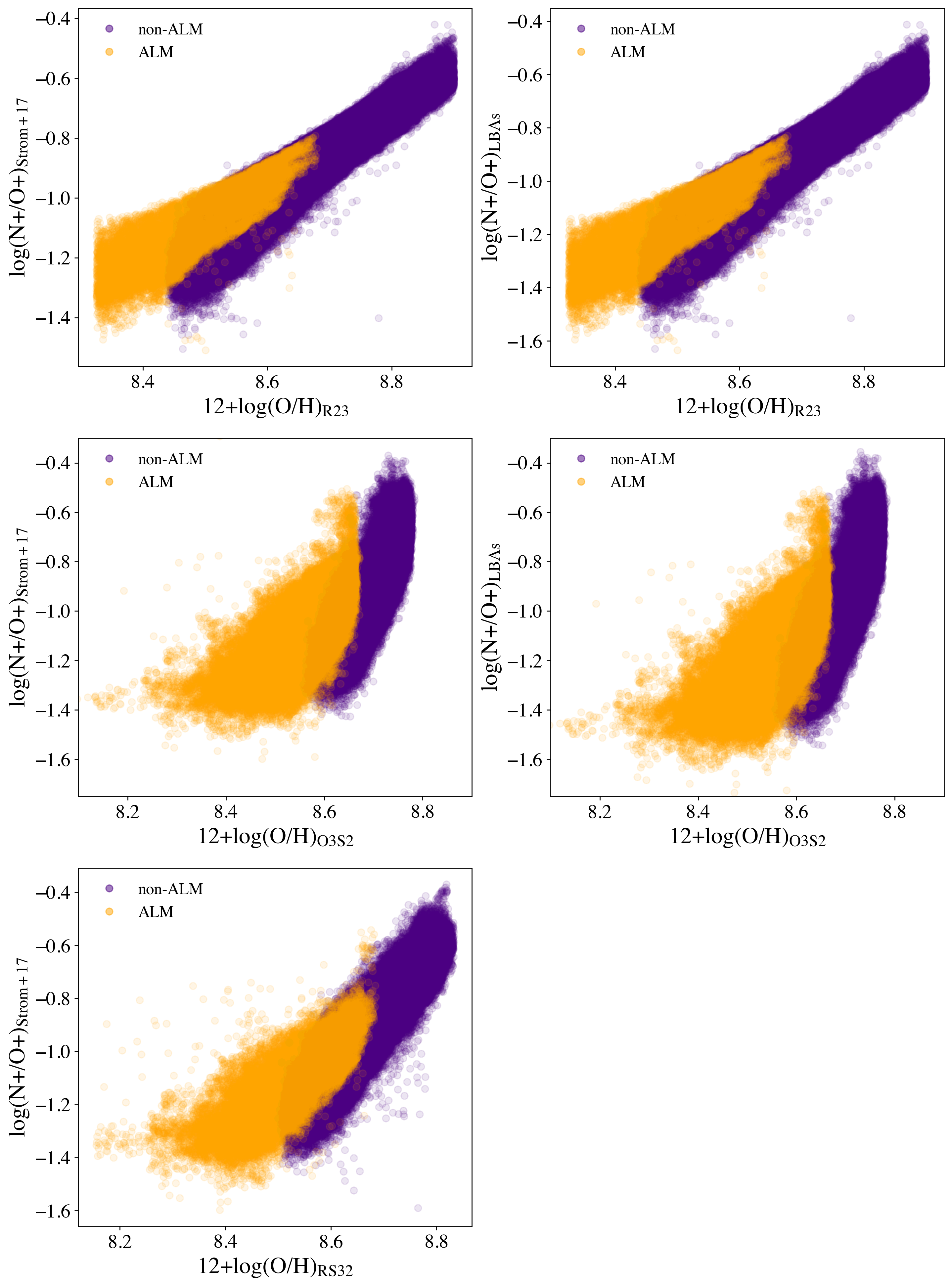}
        \caption{
                \label{fig:cali_test} 
               ALM and non-ALM spaxels plotted on the N/O vs.\ O/H plane. All combinations of strong-line methods show ALM spaxels displaced to higher N/O at fixed O/H. This means our results as demonstrated in section \ref{subsec:nooh} is robust under different choices of the metallicity calibrators.
        }
\end{figure}

\section{N/O vs.\ O/H plot with other metallicity calibrators}
To show that our results are robust under different choices of metallicity calibrators, we repeat our analysis and present the N/O vs.\ O/H plot (Fig. \ref{fig:cali_test}) with combinations of some different metallicity calibrators: O3S2, R23, RS32 in \citet{Curti_2020} for O/H, and equation (5) in \citet{Strom_2017}, Eq. \ref{eq:no} in section \ref{sec:3.2} \citep[][labeled as LBAs for Lyman break analogs]{AOH_2020} for N/O. We find all combinations of strong-line methods show ALM spaxels displaced to higher N/O at fixed O/H. This means our results as demonstrated in section \ref{subsec:nooh} in robust under different choices of the metallicity calibrators.

\bibliography{sample63}{}
\bibliographystyle{aasjournal}



\end{document}